\documentclass[twocolumn,showpacs,prb,amsfonts,amsmath,amssymb,floatfix,groupedaddress]{revtex4-1}
\usepackage{color}
\usepackage{hhline}
\usepackage{mathrsfs}
\usepackage{graphicx}
\usepackage{dcolumn}
\usepackage{bm}
\usepackage{multirow}
\usepackage{booktabs}
\usepackage{afterpage}
\usepackage{amsmath}
\usepackage{ulem}

\begin{document}

\title{Chemical physics of superconductivity in layered yttrium carbide halides from first principles}

\author{Ryosuke Akashi$^{1}$}
\thanks{ryosuke.akashi@phys.s.u-tokyo.ac.jp}
\author{Ryotaro Arita$^{2,3}$}
\author{Chao Zhang$^{4}$}
\author{K. Tanaka$^{5}$}
\author{J. S. Tse$^{5}$}
\affiliation{$^1$Department of Physics, The University of Tokyo, Hongo, Bunkyo-ku, Tokyo 113-0033, Japan}
\affiliation{$^2$RIKEN Center for Emergent Matter Science, Wako, Saitama 351-0198, Japan}
\affiliation{$^3$Departmemt of Applied Physics, The University of Tokyo, Hongo, Bunkyo-ku, Tokyo, 113-8656, Japan}
\affiliation{$^4$ Department of Physics, Yantai University, Yantai 264005, China}
\affiliation{$^5$ Department of Physics and Engineering Physics, University of Saskatchewan, 116 Science Place, Saskatoon, Saskatchewan, S7N 5E2, Canada}

\date{\today}
\begin{abstract}
We perform a thorough first-principles study on superconductivity in yttrium carbide halide Y$_{2}$$X_{2}$C$_{2}$ ($X$=Cl, Br, I) whose maximum transition temperature ($T_{\rm c}$) amounts to $\sim$10~K. A detailed analysis on the optimized crystal structures reveals that the Y$_{2}$C$_{2}$ blocks are compressed uniaxially upon the halogen substitution from Cl, Br to I, contrary to the monotonic expansion of the lattice vectors. With a nonempirical method based on the density functional theory for superconductors within the conventional phonon mechanism, we successfully reproduce the halogen dependence of $T_{\rm c}$. Anomalously enhanced coupling of one C$_{2}$ libration mode is observed in Y$_{2}$I$_{2}$C$_{2}$, which imply possible departure from the conventional pairing picture. Utilizing the Wannier representation of the electron-phonon coupling, we show that the halogen electronic orbitals and ionic vibrations scarcely contribute to the superconducting pairing. The halogen dependence of this system is hence an indirect effect of the halogen ions through the uniaxial compressive force on the superconducting Y$_{2}$C$_{2}$ blocks. We thus establish a quantitatively reliable picture of the superconducting physics of this system, extracting a unique effect of the atomic substitution which is potentially applicable to other superconductors.
\end{abstract}
\pacs{}

\maketitle

\section{Introduction}
Synthesizing high-temperature superconductors by design has been a long-sought goal of materials physics and chemistry. Investigating light element compounds is a rational strategy derived from the Bardeen-Cooper-Schrieffer theory of phonon-mediated superconductivity,~\cite{BCS1,BCS2} where the transition temperature $T_{\rm c}$ is proportional to the frequency of phonons mediating the electron pairing. The recent predictions~\cite{Li2014,Duan2014,LaH10-Hemley-PNAS2017} and later (or almost concurrent) discoveries~\cite{Eremets,LaH10-Hemley-PRL2019,LaH10-Eremets-Nature2020} of hydride superconductors under pressure is a milestone of the theory, which was finally brought about by the advance of the first-principles electronic structure calculation and crystal structure prediction methods.~\cite{Oganov-book, Flores-Livas2020} 

In this context, carbon compounds have also long been a subject of intense study. Owing to the relatively small mass and strong covalent bonding of carbon atoms, they may host high-frequency phonons. Among them, the rare-earth carbides $Re_{2}$C$_{3}$~ ($Re=$rare earth) (Refs.~\onlinecite{La2C3-1969,Y-Th-C-1969,Y2C3-2004}), $Re$C$_{2}$~(Refs.~\onlinecite{YC2-1968,ReC2-1997}) and carbide halide $Re_{2}$$X_{2}$C$_{2}$~(Refs.~\onlinecite{Simon-ZAAC1996,Henn-YXC-PRL}) are interesting materials. They generally show superconductivity at temperatures around $\sim$10~K; especially in the sesquicarbide $T_{\rm c}$ reaches 18~K at ambient pressure.~\cite{Y2C3-2004} Remarkably, they contain carbon dimers in their crystal structures. From the physics viewpoint these materials are attractive as they can be model systems where $T_{\rm c}$ is boosted by the high frequency vibrations of the carbon dimers. On the other hand, from the chemistry viewpoint, their formula units are significant since the valence electrons cannot be straightforwardly attributed to the constituent atoms from their regular ionic valence.  According to the Zintl-Klemm concept each carbon dimer should host additional valence electrons and exist as a molecular anion (C$_{2}$)$^{-n}$. Inspired by this characteristics, a ``chemical" theory of superconductivity for this system has been proposed: The superconducting pairing occurs locally at the dimer sites, where extra electrons are confined as a consequence of the Zintl-Klemm rule.~\cite{Simon-review} Superconductivity in the rare-earth carbides thus serves an interesting intersection of physics and chemistry.

In this work, we conduct a first-principles study of the electronic, phononic and superconducting properties of the yttrium carbide halide Y$_{2}$$X_{2}$C$_{2}$ ($X=$Cl, Br, I) to establish its chemical-physical description. This system becomes superconducting with $T_{\rm c}\sim$11.6~K at maximum.~\cite{Simon-ZAAC1996,Henn-BrI-2001,Ahn-pressure-JPhysC2016} A distinguished feature of this system is its two-dimensional layered crystal structure, where the yttrium blocks containing the carbon dimers and intercalant halogen blocks are alternately stacked. Substitution of the halogen atoms yields systematic change of $T_{\rm c}$ from $\sim$2.5~K ($X=$Cl) to 5.0 (Br) and 10~K (I).~\cite{Simon-ZAAC1996} The electron-phonon coupling properties in some related materials such as Y$_{2}$C$_{3}$ (Ref.~\onlinecite{Singh-Mazin-Y2C3}) and YC$_{2}$ (Ref.~\onlinecite{XUE2019120}) have been studied with the first-principles calculations, and found that the mechanism conforms to the conventional phonon-mediated superconducting theory. For the carbide halide, calculations of its electronic band structure and Raman-active modes have been reported,~\cite{Puschnig-PRB2001} but the electron-phonon and superconducting calculations are yet unprecedented. Also, a perspective on the halogen dependence is lacking since the study on the low-$T_{\rm c}$ chloride is scarce. Actually the carbon isotope effect in this system is not as evident as in the other yttrium carbides:~\cite{Schnelle-isotope-JAP1998} The two-dimensional structure specific to the carbide halide may induce some anomaly beyond the conventional phonon pairing mechanism.

We report thorough first-principles calculations of the electron-phonon coupling to examine whether the experimentally observed superconducting transition temperatures can be reasonably explained within the conventional phonon-mediated scenario. From the optimized crystal structures, we find an effect of the halogen ions as a source of local pressure on the superconducting Y$_{2}$C$_{2}$ layers. Using the Wannier representation of the electron-phonon coupling matrix elements, we disentangle the interplay of the atomic orbitals, through which a chemical view on the superconducting electronic states is developed. The pioneering local Zintl-Klemm pairing theory~\cite{Simon-review} is finally found to be not as relevant as expected. The superconducting transition temperatures are evaluated non-empirically with the density functional theory for superconductors, with which the halogen atom dependence of $T_{\rm c}$ is well reproduced. Possible departure from the conventional pairing scenario is also suggested in Y$_{2}$I$_{2}$C$_{2}$, which exhibit the highest $T_{\rm c}$ among the stoichiometric Y$_{2}$X$_{2}$C$_{2}$ series. 

\section{Method and computational details}
We calculated the electronic and phononic properties of Y$_2$$X_2$C$_2$ ($X=$Cl, Br, I) from first principles with the plane-wave pseudopotential method implemented in {\sc Quantum Espresso}.~\cite{QE} We solved the normal-state Kohn-Sham equation~\cite{Hohenberg-Kohn1964,Kohn-Sham1965} for the electronic normal state and later analysis of the phonon and superconducting states. The Troullier-Martins norm-conserving pseudopotentials~\cite{TM} for the Y, C and halogen atoms were employed. The Perdew-Burke-Ernzerhof generalized gradient approximation for solids (PBEsol)~\cite{PBEsol} was used for the exchange-correlation functional. The phononic properties were calculated with the density functional perturbation theory~\cite{DFPT-review} as implemented in {\sc ph.x} package.

We examine the pairing characteristics within the conventional Eliashberg theory~\cite{Eliashberg1960,Scalapino-bookchap, Schrieffer-book} of phonon-mediated superconductivity with the Migdal approximation.~\cite{Migdal1958} We calculate the Eliashberg spectral function $\alpha^{2}F(\omega)$,\cite{Scalapino-bookchap} which represents the electron-phonon coupling strength for the pairing:
\begin{eqnarray}
\alpha^2F(\omega)
&=&
\frac{1}{N(0)}\sum_{{\bf q}\nu} \sum_{nn'{\bf k}} |g^{\nu {\bf q}}_{n{\bf k}+{\bf q}, n'{\bf k}}|^2 \delta(\omega-\omega_{{\bf q}\nu}) \nonumber \\
&& \hspace{40pt}\times \delta(\xi_{n{\bf k}+{\bf q}}) \delta(\xi_{n'{\bf k}})
.
\label{eq:alpha2F}
\end{eqnarray}
The electronic Kohn-Sham and phononic vibrational eigenstates are respectively labeled by the band index $n$ ($n'$), electronic wave vector ${\bf k}$, mode index $\nu$ and phonon wave vector ${\bf q}$. $\xi_{n{\bf k}}$ denotes the normal-state Kohn-Sham energy eigenvalue measured from the Fermi level. $\omega_{\nu {\bf q}}$ is the phonon frequency. $N(\xi)$ denotes the electronic density of states (DOS). The electron-phonon coupling coefficient $g^{\nu {\bf q}}_{n{\bf k}+{\bf q}, n'{\bf k}}$ is defined as the matrix element between the Kohn-Sham states
\begin{eqnarray}
g^{\nu {\bf q}}_{n{\bf k}+{\bf q}, n'{\bf k}}
=
\langle n{\bf k}+{\bf q}| \partial_{\nu {\bf q}} V_{\rm KS} |n'{\bf k}\rangle
,
\end{eqnarray}
where $\partial_{\nu {\bf q}} V_{\rm KS}$ denotes the derivative of the self-consistent Kohn-Sham potential with respect to the atomic positions in the direction of mode displacement vector ${\bf u}_{{\bf q}\kappa}^{\nu}$, with $\kappa\equiv(\alpha, i)$ being the abbreviated index of atom $\alpha$ and direction $i=x, y, z$. We also evaluated the parameters 
\begin{eqnarray}
\lambda=2\int d\omega \frac{\alpha^2 F(\omega)}{\omega},
\label{eq:lambda}
\\
\omega_{\rm ln}
=
{\rm exp}
\left[
\frac{2}{\lambda}
\int d\omega
\frac{\alpha^2 F(\omega)}{\omega}
{\rm ln}\omega
\right],
\\
\bar{\omega}_{2}
=
\left[
\frac{2}{\lambda}
\int d\omega
\frac{\alpha^2 F(\omega)}{\omega}
\omega^2
\right]^{\frac{1}{2}},
\end{eqnarray}
which enter the McMillan-Allen-Dynes semiempirical formula for the superconducting transition temperature,~\cite{Allen-Dynes}
\begin{eqnarray}
T_{\rm c}^{\rm McM}
=f_{1}f_{2}\frac{\omega_{\rm ln}}{1.2}
{\rm exp}
\left[
-\frac{1.04(1+\lambda)}{\lambda -\mu^{\ast}(1+0.62\lambda)}
\right]
.
\label{eq:McMillan}
\end{eqnarray}
The Coulomb pseudopotential $\mu^{\ast}$ represents the renormalized pair-breaking Coulomb repulsion.~\cite{Morel-Anderson, McMillan} The correction factors $f_{1}=f_{1}(\lambda, \mu^{\ast})$ and $f_2=f_2(\lambda, \omega_{\rm ln},\bar{\omega}_{2}, \mu^{\ast})$, which are both approximately unity in the weak-coupling limit, are defined in Ref.~\onlinecite{Allen-Dynes}. Finally we calculated the area under $\alpha^{2}F(\omega)$, $A$, which may better correlate with $T_{\rm c}$ than $\lambda$ or $\omega_{\rm ln}$ alone,~\cite{Leavens-Carbotte1974,Carbotte-RMP1990,Quan-Pickett2019}
\begin{eqnarray}
A=
\int d\omega \alpha^2 F(\omega)
.
\end{eqnarray}

We adopted the Wannier representation of the electron-phonon coupling coefficient~\cite{Wannier-elphPRB2007} for two purposes. First, we executed the efficient interpolation of $g^{\nu {\bf q}}_{n{\bf k}+{\bf q}, n'{\bf k}}$ on dense ${\bf k}$ and ${\bf q}$ meshes for convergence of the summations $\sum_{{\bf q}\nu} \sum_{nn'{\bf k}}$ in Eq.~(\ref{eq:alpha2F}) using the inverse and forward Fourier transformations as formulated in Ref.~\onlinecite{Wannier-elphPRB2007}. Second, we performed decomposition of $\alpha^2F(\omega)$ into local electronic and phononic contributions. Namely, 
\begin{eqnarray}
\alpha^2F(\omega)
&=&
\frac{1}{N(0)}\sum_{{\bf q}\nu} \sum_{nn'{\bf k}} |g^{\nu {\bf q}}_{n{\bf k}+{\bf q}, n'{\bf k}}|^2 \delta(\omega-\omega_{{\bf q}\nu}) \nonumber \\
&& \hspace{40pt}\times \delta(\xi_{n{\bf k}+{\bf q}}) \delta(\xi_{n'{\bf k}})
\\
&=& \frac{1}{N(0)}\sum_{{\bf q}\nu} \sum_{nn'{\bf k}} \sum_{\kappa \kappa'}  \sum_{\substack{m_{1}m_{2} \\ m_{3}m_{4}}} G_{nn'; m_{1}m_{2} m_{3}m_{4}}^{\nu; \kappa \kappa'}({\bf k}, {\bf q})  \nonumber \\
&& \hspace{40pt}\times \delta(\omega-\omega_{{\bf q}\nu}) \delta(\xi_{n{\bf k}+{\bf q}}) \delta(\xi_{n'{\bf k}})
.
\label{eq:a2F-tot}
\end{eqnarray}
Indices $m_{1}$---$m_{4}$ run over the Wannier orbitals. Factor $G_{nn'; m_{1}m_{2} m_{3}m_{4}}^{\nu; \kappa \kappa'}({\bf k}, {\bf q})$ is defined as
\begin{eqnarray}
G_{nn';m_{1}m_{2} m_{3}m_{4}}^{\nu; \kappa \kappa'} ({\bf k}, {\bf q})
\!&\equiv&\!
[u^{\nu}_{{\bf q} \kappa} \tilde{g}^{\kappa {\bf q}}_{m_{1}{\bf k}+{\bf q}, m_{2}{\bf k}} 
U_{nm_{1},{\bf k}+{\bf q}} U^{\ast}_{n'm_{2},{\bf k}}]^{\ast} 
\nonumber \\
&&
\hspace{-20pt}
\times
u^{\nu}_{{\bf q} \kappa'}\tilde{g}^{\kappa' {\bf q}}_{m_{3}{\bf k}+{\bf q}, m_{4}{\bf k}} 
U_{nm_{3},{\bf k}+{\bf q}} U^{\ast}_{n'm_{4},{\bf k}}
.
\label{eq:Gfunc}
\end{eqnarray}
Here, $u^{\nu}_{{\bf q} \kappa}$ denotes the displacement vector of mode $\nu {\bf q}$ with weights in inverse proportion to the square root of the atomic mass values and $U_{nm_{1},{\bf k}}$ denotes the unitary transformation between the Bloch to Wannier space, as defined in Ref.~\onlinecite{Wannier-elphPRB2007}. Coefficient $\tilde{g}^{\kappa{\bf q}}_{m{\bf k}+{\bf q}, m'{\bf k}} $ thus means the matrix element between the Bloch sums of the individual Wannier orbitals $m$ and $m'$ via the collective displacements of $\kappa$ with wave vector ${\bf q}$. The unambiguous decompositions of the Eliashberg function is then defined from Eqs.~(\ref{eq:a2F-tot}) and (\ref{eq:Gfunc}) as
\begin{eqnarray}
\alpha^2F(\omega)
&=&\sum_{\kappa \kappa'} \sum_{\substack{m_{1}m_{2} \\ m_{3}m_{4}}} \alpha^{2}F^{\kappa\kappa'}_{m_{1}m_{2} m_{3}m_{4}} (\omega)
\label{eq:a2F-decomp}
\end{eqnarray}

In this study, we calculated the electronic Wannier orbitals using a preliminary version of {\sc RESPACK}~\cite{PhysRevB.93.085124,doi:10.1143/JPSJ.78.083710,doi:10.1143/JPSJ.77.093711,PhysRevB.79.195110,doi:10.1143/JPSJ.72.777,RESPACK-arxiv} code, which is the implementation of the method proposed by Souza, Marzari and Vanderbilt for entangled energy bands.~\cite{MLWF1, MLWF2}

We perform calculations of the superconducting transition temperature $T_{\rm c}$ in two ways within the Eliashberg theory~\cite{Eliashberg1960, Scalapino-bookchap,Schrieffer-book} with the Migdal approximation.~\cite{Migdal1958} The first is the isotropic Eliashberg equations,
\begin{eqnarray}
Z(i\omega_{n})
&=&
1+\frac{\pi T}{\omega_{n}}\sum'_{m}\frac{\omega_{m}}{R(i\omega_{m})}
\lambda(\omega_{n}-\omega_{m})
,
\label{eq:Iso-Eliashberg1}
\\
\Delta(i\omega_{n})
&=&
\frac{\pi T}{Z(i\omega_{n})}\sum'_{m}\frac{\Delta(i\omega_{m})}{R(i\omega_{m})}
\left[
\lambda(\omega_{n}-\omega_{m})
-\mu^{\ast}
\right],
\nonumber \\
\label{eq:Iso-Eliashberg2} \\
R(i\omega_{m})
&=&
\sqrt{\omega_{m}^2 + \Delta^2(i\omega_{m})}
,
\\
\lambda(\omega_{n}-\omega_{m})
&=&
2\int_{0}^{\infty}d\omega \frac{\omega\alpha^2 F(\omega)}{\omega^2+(\omega_{n}-\omega_{m})^2},
\end{eqnarray}
where $\omega_{n}$ denotes the fermionic Matsubara frequency. Sum $\sum'_{m}$ is limited by the cutoff $|\omega_{m}|\leq\omega_{\rm max}$. The effect of the renormalized Coulomb repulsion is represented by the parameter $\mu^{\ast}$, which is conceptually a function of $\omega_{\rm max}$ but treated as an adjustable parameter. The problem of calculating the superconducting transition temperature from these equations is recast to the following eigenvalue problem with a transformed order parameter $\bar{\Delta}$~(Ref.~\onlinecite{Tanaka-Tse2009}):
\begin{eqnarray}
\rho \bar{\Delta}(i\omega_{n})
=
T\sum'_{m}
\left\{
\lambda(\omega_{n}-\omega_{m})
-\mu^{\ast}
-\delta_{mn}\frac{|\tilde{\omega}_{m}|}{\pi T}
\right\}\bar{\Delta}(i\omega_{m}) 
\nonumber \\
\label{eq:Iso-Eliashberg3}
\end{eqnarray}
with
\begin{eqnarray}
\tilde{\omega}_{n}
=
\omega_{n}
+\pi T \sum'_{m}{\rm sgn} \omega_{m}\lambda(\omega_{n}-\omega_{m})
.
\end{eqnarray}
$T_{\rm c}$ is then given by the condition $\rho(T_{\rm c})=0$. The functional derivative $\delta T_{\rm c}/\delta \alpha^2 F(\omega)$, showing at which frequency the coupling effect on $T_{\rm c}$ is most significant, is evaluated using $\rho(T)$ and $\bar{\Delta}(i\omega_{n})$ by the Bergmann-Rainer formula.~\cite{Bergmann-Rainer, Tanaka-Tse2009}

Finally, we evaluate the superconducting transition temperature with the gap equation from the density functional theory for superconductors,~\cite{SCDFTI, SCDFTII}
\begin{equation}
	\Delta_{n\bm{k}} = -\mathcal Z_{n\bm{k}}\Delta_{n\bm{k}}
	- \frac{1}{2}\sum_{n'\bm{k'}}\mathcal K_{n\bm{k}n'\bm{k'}}
	\frac{\tanh[(\beta/2)E_{n'\bm{k'}}]}{E_{n'\bm{k'}}}\Delta_{n'\bm{k'}}.
	\label{eq:SCDFTgap}
\end{equation}
Here, $\beta$ is the inverse temperature, $\Delta_{n{\bf k}}$ is the gap function and $E_{n\bm k}$ is defined as $E_{n\bm k}=\sqrt{\xi_{n\bm k}^2 + |\Delta_{n\bm k}|^2}$. The exchange-correlation kernels $\mathcal{K}=\mathcal{K}^{\rm ph}+\mathcal{K}^{\rm el}$ and $\mathcal{Z}=\mathcal{Z}^{\rm ph}$ quantitatively treats the interaction effects within the Migdal-Eliashberg theory. Notably, the retardation effect on the effective Coulomb repulsion is treated without any empirical parameter like $\mu^{\ast}$. It is achieved approximately, instead of using the Matsubara frequency, with the Kohn-Sham energy dependences of $\mathcal{K}^{\rm ph}$ and $\mathcal{K}^{\rm el}$.~\cite{SCDFTI, SCDFTII} The $n{\bf k}$-averaged forms for $\mathcal{K}^{\rm ph}$ [Eq.~(23) in Ref.~\onlinecite{SCDFTII}] and $\mathcal{Z}^{\rm ph}$ [Eq.~(40) in Ref.~\onlinecite{Akashi-Z-asym}] were employed for the phononic terms, which can be calculated using the Eliashberg function $\alpha^{2}F(\omega)$. The electronic term $\mathcal{K}^{\rm el}$, which describes the screened Coulomb repulsion, was treated with the fully $n{\bf k}$-dependent form: we adopted the formula of Eq.~(3) in Ref.~\onlinecite{Akashi-H2S-SCDFT}, in which the dielectric matrix within the random phase approximation is used for the plasmonic effect.\cite{Takada1978,Akashi-plasmon-PRL2013,Akashi-plasmon-JPSJ2014}. Those terms were evaluated using the normal-state properties as formulated in the above references.



In Table~\ref{tab:conditions} we summarize the detail conditions of the electron, phonon and superconducting calculations. We consistently employed the tetrahedron interpolation methods for the singular integrals such as Eq.~(\ref{eq:alpha2F}), which, in contrast to the smearing methods, assure convergence to the exact integrals with efficient number of wave-number grid points.

\begin{table*}[h!]
\caption[t]
{Detailed settings for the calculations. The format of the ${\bf k}$ and ${\bf q}$ point grids obeys the convention of {\sc Quantum Espresso}, where the second lattice vector is taken to be the interlayer one. Subscript ``1" for {\bf q} points denotes the mesh with displacement by half a grid step in the first and third directions for avoiding divergence at ${\bf q}$=$\Gamma$. The wave function and charge density cutoffs for the charge density and dynamical matrix calculations were set to 80~Ry and 320~Ry, respectively.}
\begin{center}
\label{tab:conditions}
\tabcolsep = 1mm
\begin{tabular}{|l |c|c|} \hline
 &&Method or setting  \\ \hline
charge density &{\bf k} & (12 4 12)  \\ \cline{2-3}
                      &interpol. &Optimized tetrahedron~\cite{opt-tetra} \\ \hline
dynamical matrix &{\bf k} & (12 4 12) \\ \cline{2-3}
                        &{\bf q} & (6 2 6)  \\ \cline{2-3}
                        &interpol. &Optimized tetrahedron~\cite{opt-tetra} \\ \hline
Wannier functions &{\bf k} & (6 2 6) \\ 
 &Initial guess of orbitals& Y-$d$, $X$-$p$ and C-$sp$ \\ 
 &Outer window (eV) & [-18.7, 21.3]\\
 &Inner (frozen) window (eV) & [-1.0, 1.0] \\\hline 
Eliashberg function &{\bf k}$^{\dagger}$ & (36 12 36)  \\ \cline{2-3}
                           &{\bf q}$^{\dagger}$ & (18 6 18)$_{1}$ \\ \cline{2-3}
                        &interpol. &Optimized tetrahedron~\cite{opt-tetra} \\ \hline
dielectric function &{\bf k}  for bands crossing $E_{\rm F}^{\dagger\dagger}$ & (18 6 18)  \\ \cline{2-3}
                           &{\bf k}  for other bands& (6 2 6)  \\ \cline{2-3}
                           &{\bf q}  &  (6 2 6) \\ \cline{2-3}
                           &unoccupied band num.& $\sim$227 \\ \cline{2-3}
                           &interpol. &Tetrahedron with the Rath-Freeman treatment~\cite{Rath-Freeman}  \\ \cline{2-3}
                           &Plane wave cutoff& 12.8~Ry  \\ \hline
DOS for phononic kernels &{\bf k}   & (27 9 27)  \\ \cline{2-3}
                           &interpol. &Tetrahedron with the Bl\"{o}chl correction~\cite{Bloechl-tetra} \\ \hline
Eliashberg equations solution & Matsubara frequency cutoff $\omega_{\rm max}$ & 6$\times$ $\omega_{\alpha^{2}F}\ ^{\dagger\dagger\dagger}$ \\ \hline
SCDFT gap function & unoccupied band num. &107 \\ \cline{2-3} 
           &{\bf k} for the electronic kernel & (6 2 6)  \\ \cline{2-3} 
           &{\bf k} for the KS energy eigenvalues & (27 9 27)  \\ \cline{2-3} 
           & $N_{\rm s}$ for bands crossing $E_{\rm F}$ &6000 \\ \cline{2-3} 
           & $N_{\rm s}$ for other bands &200\\ \cline{2-3}
           & Sampling error in $T_{\rm c}$ &$\lesssim$7\% \\ \hline 
           \multicolumn{3}{l}{$^\dagger$ Electron and phonon energy eigenvalues and matrix elements were calculated on these auxiliary grid points.} \\
           \multicolumn{3}{l}{$^{\dagger\dagger}$ Electron energy eigenvalues were calculated on these auxiliary grid points.} \\
           \multicolumn{3}{l}{$^{\dagger\dagger\dagger}$ $\omega_{\alpha^{2}F}$ was taken to be the maximum phonon frequency below which the calculated $\alpha^{2}F(\omega)$ values are nonzero.} \\
\end{tabular}
\end{center}
\end{table*}

\clearpage

\begin{figure}[h!]
 \begin{center}
  \includegraphics[scale=0.45]{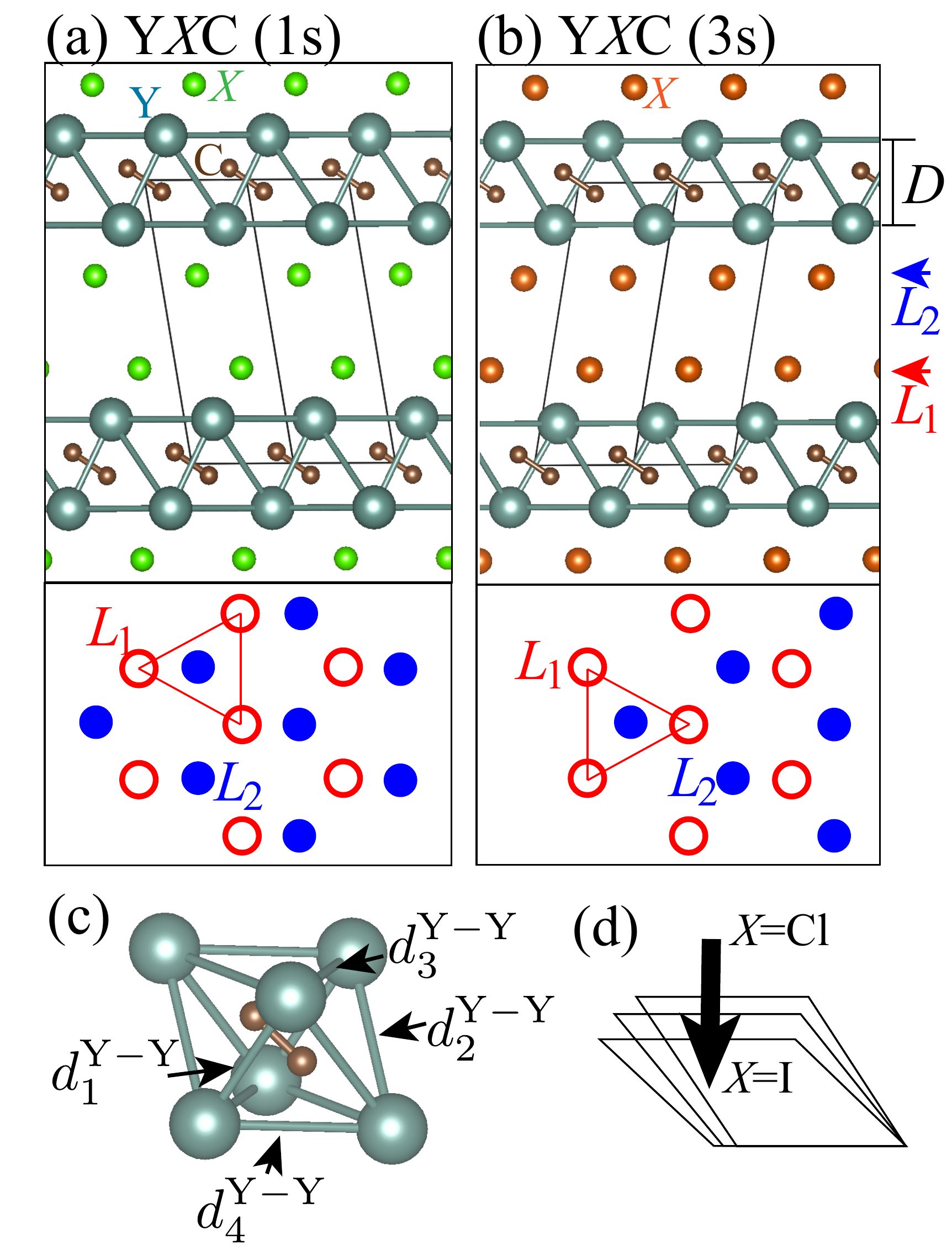}
  \caption{Crystal structures of Y$_{2}$$X_{2}$C$_{2}$ with (a) 1s and (b) 3s stacking geometries. The definition of the yttrium interplane distance $D$ is indicated. In the lower panels the top views of the neighboring halogen layers $L_{1}$ and $L_{2}$ are shown. (c) Yttrium octahedron containing the carbon dimer, the array of which forms the individual Y$_2$C$_2$ block layers. (d) Schematic of the Poisson-type structural change of the yttrium framework by the halogen substitution, where the degree of changes is exaggerated.} 
  \label{fig:cryst}
 \end{center}
\end{figure}

\begin{table}[t]
\caption[t]
{Optimized lattice parameters compared with the experimentally observed values.~\cite{Simon-ZAAC1996}}
\begin{center}
\label{tab:struct}
\tabcolsep = 1mm
\begin{tabular}{lcccccc} \hline
&\multicolumn{2}{c}{YClC} & \multicolumn{2}{c}{YBrC} & \multicolumn{2}{c}{YIC}  \\ 
& Calc. & Expt. & Calc. & Expt. &Calc. & Expt. \\ \hline
$a$ (\AA) &6.761& 6.820  &6.872 &6.953  &7.114&7.212  \\ 
$b$ (\AA) &3.672 & 3.713 &3.718 & 3.764 &3.809 &3.876  \\ 
$c$ (\AA) & 9.453 & 9.327 &9.914 & 9.938 & 10.205 &10.411 \\ 
$\beta$ (deg) &95.31 &94.75 & 99.49 &99.98  &93.80 &93.55  \\  \hline
\end{tabular}
\end{center}
\end{table}

\begin{table*}[t!]
\caption[t]
{Optimized bond lengths characterizing the yttrium octahedron containing the carbon dimer compared with the experimentally observed values.~\cite{Simon-ZAAC1996} The second column shows the degree of degeneracy of the respective bonds. The interplane distances $D$ as defined in Fig.~\ref{fig:cryst} are also shown. All the values are in units of \AA.}
\begin{center}
\label{tab:struct2}
\tabcolsep = 1mm
\begin{tabular}{lccccccc} \hline
\multicolumn{2}{c}{  } &\multicolumn{1}{c}{$X=$Cl} & \multicolumn{2}{c}{$X=$Br} & \multicolumn{2}{c}{$X=$I}  \\ 
 &deg.& Calc. & Calc. & Expt. &Calc. & Expt. \\ \hline
$d^{{\rm Y-Y}}_{1}$  & 1 & 3.372 &3.356&3.400(6) &3.319&3.370(7)  \\ 
$d^{{\rm Y-Y}}_{2}$ & 2 &3.899  &3.854 & \multirow{2}{*}{3.764(1), 3.852(5)}  &3.771 &  \multirow{2}{*}{3.726(6), 3.866(2)} \\ 
$d^{{\rm Y-Y}}_{3}$ & 2 & 3.675  &3.719 &  &3.811 &   \\ 
$d^{{\rm Y-Y}}_{4}$ & 4 &3.847 &3.907 & 3.953 &4.035 &4.075(3)  \\  
Average & & 3.767&3.792 & 3.827& 3.847&3.872 \\ \hline

$d^{{\rm C-C}}$ & &  1.330&1.327&1.267(12) &1.322&1.304(6)  \\ \hline
$D$ (\AA) & & 2.955 &2.895& &2.763 & \\ \hline
\end{tabular}
\end{center}
\end{table*}

\section{Results}
\subsection{Structural properties}
\label{sec:struct}
The series of Y$_2$$X_2$C$_2$ compounds exhibit base-centered monoclinic lattice for all the cases $X$=Cl, Br, I. At ambient pressure, they crystallize in two slightly different stacking forms $1s$ and $3s$, belonging to the same space group $C2/m$ (Fig.~\ref{fig:cryst}). These structures have the common $X$-Y-C$_2$-Y-$X$ layers as the building blocks and related by changing the interlayer latice vectors. Their naming convention is based on the approximate periods in the vertical direction of some representative systems.~\cite{Simon-ZAAC1996} We comply with this convention in this paper, although we point out that the two types of stacking are more definitely distinguished by the positions of the neighboring halogen layers [Fig.~\ref{fig:cryst}(a)(b)], over which of the two triangle units the upper halogen atoms are located. At ambient pressure the $X$=Cl and I systems take the $1s$ stacking whereas the $X$=Br system takes the $3s$ stacking.~\cite{Simon-ZAAC1996} A previous first principles simulation indicates that the formation energies of $1s$ and $3s$ stacking are very close.~\cite{Ahn-pressure-JPhysC2016} We adopted the experimentally observed stacking forms as the initial conditions for the structure optimization. The optimized lattice parameters are summarized in Table~\ref{tab:struct}. We also list the optimized bonding lengths of the yttrium octahedra containing the carbon dimers (Fig.~\ref{fig:cryst}(c)) in Table~\ref{tab:struct2}. The values show fair agreement with the experimental data. A systematic underestimation of the structural parameters by $\sim$2~\% was observed, though it does not affect the discussion in this section.

As the halogen atoms are substituted from the lighter (Cl) to heavier (I), the optimized lattice vectors tend to become longer, especially in the interlayer direction $c$. Although this trend is a straightforward consequence of the difference in the halogen ionic radii, we also find its remarkable effect on the Y$_{2}$C$_{2}$ layered blocks. The distance between the upper and lower yttrium planes indicated as $D$ in Fig.~\ref{fig:cryst} decreases by 7\% with substitution from Cl to I. This trend is contrary to the expansion of the lattice vectors. Previously, this unusual observation was attributed to the local distortion of the yttrium octahedra.~\cite{Puschnig-PRB2001} Here we suggest an alternative as a uniaxial compression and lateral expansion of the Y$_{2}$C$_{2}$ blocks [Fig.~\ref{fig:cryst}(d)]. It is, in fact, a microscopic Poisson effect where the halogen atoms exert on the blocks uniaxial pressure whose strength depends on their ionic radii. This perspective has apparently been obscured by the expansion of the lattice vectors. It is hence encouraged to revisit the experimental~\cite{Simon-ZAAC1996,Henn-neutron-PRB1998} and previous first-principles~\cite{Puschnig-PRB2001} data to see if this Poisson-type structural change of the yttrium framework is indeed presented.



\begin{figure*}[t!]
 \begin{center}
  \includegraphics[scale=0.28]{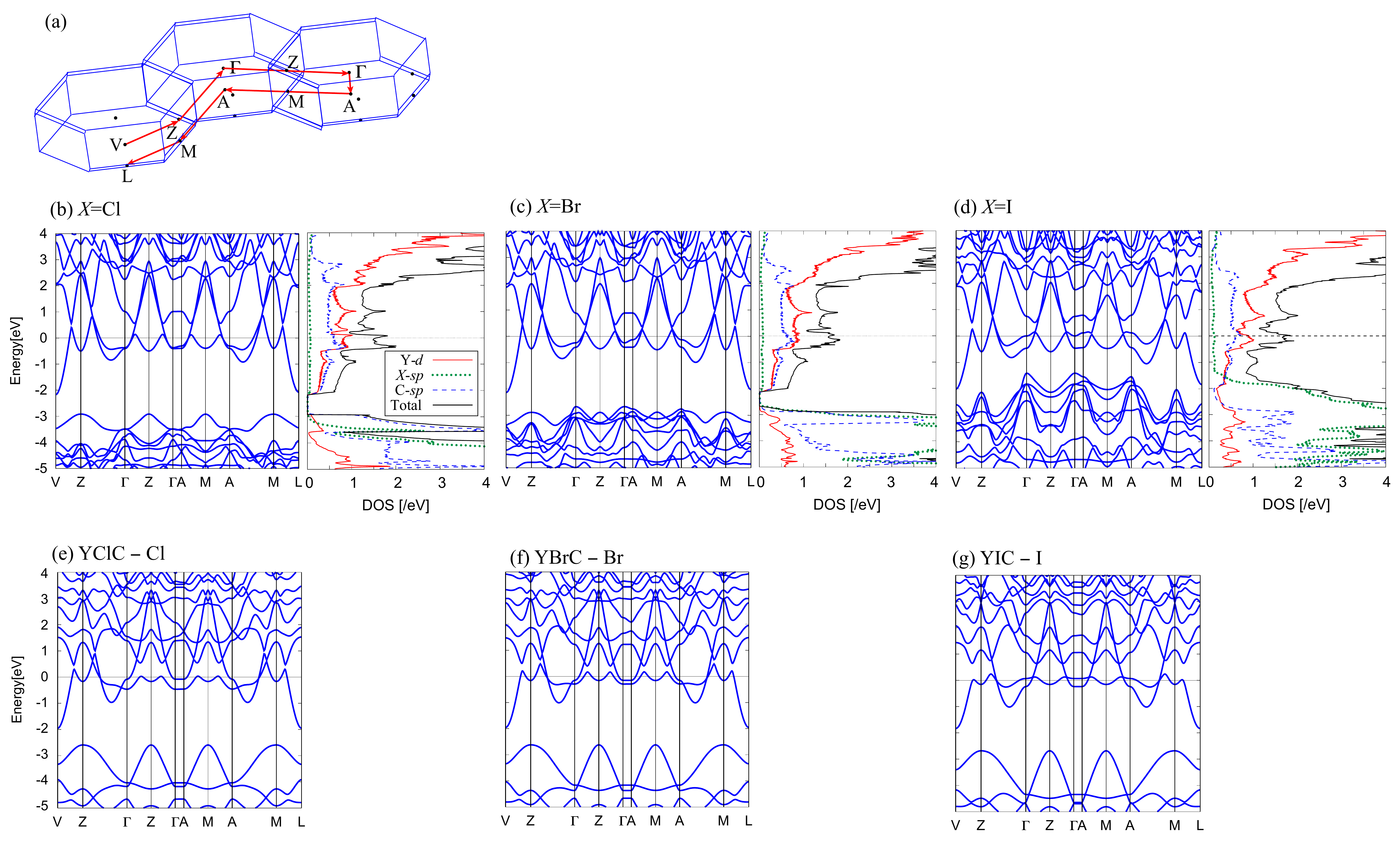}
  \caption{Elecronic band structures of Y$_{2}$$X_{2}$C$_{2}$. (a) The Brillouin zone and ${\bf k}$-point path. (b)--(d) Band structures, total DOS and partial DOS spectra. (e)--(g) Band structures of the artificial systems Y$_{2}$$(e^{-})_{2}$C$_{2}$, whose crystal structures were constructed by subtracting the halogen atoms from the optimized structures of the original stoichiometric systems.} 
  \label{fig:bands-sum}
 \end{center}
\end{figure*}

\subsection{Electronic properties}
We next examine the halogen-atom dependence of the electronic band structure. The calculated band structures and density of states (DOS) are shown in Fig.~\ref{fig:bands-sum}. The most pronounced change in the band structure from the halogen substitution is the gap opening of the two nearly degenerate bands at the $\Gamma$ point. The partial DOS spectra derived by the projection onto the Wannier orbitals indicate that the metallic bands are formed by the carbon-$p$ and yttrium-$d$ orbitals. This is consistent with the elementary molecular-orbital model~\cite{Simon-ZAAC1996} and previous first-principles calculations using localized basis sets.~\cite{Simon-ZAAC1996,Puschnig-PRB2001} The halogen-$p$ states are consistently fully occupied, indicating an ionic character of the bonding between the halogen atoms and YC blocks. Their energy levels move upward by the substitution from the lighter to heavier ones, which reflects the change in the electron affinity. 

The effects of the halogen atoms on the band structures can be divided into three contributions. The halogen atoms (i) exert internal uniaxial pressure which distorts the YC blocks, (ii) serve as sources of the one-body ionic potential for the YC metallic states, and (iii) provide local orbitals that mediate the interlayer hybridization across the YC blocks. To examine these contributions quantitatively, we calculated the band structures in hypothetical crystals where the halogen atoms are removed and compensating uniform charges are filled instead. The remarkable resemblance of the band structures with and without the halogen atoms (upper to lower panels) shows that the metallic properties are dominated by the YC layers. We can see effect (i) on the YC layers from the changes from panels (e), (f) to (g). The bands at the $\Gamma$ points are inverted with the halogen substitution. This is closely related to the shape of the Y-octahedron. In the chloride system $d_{2}^{{\rm Y-Y}}>d_{4}^{{\rm Y-Y}}$, whereas in the iodide $d_{2}^{{\rm Y-Y}}<d_{4}^{{\rm Y-Y}}$ (see Fig.~\ref{fig:cryst}(c) for the definition of the bonds). The band inversion occurs through the crossover regime $d_{2}^{{\rm Y-Y}} \simeq d_{4}^{{\rm Y-Y}}$. From the comparison between panels (b) and (e), effect (ii) also contributes to this inversion. Effect (iii) results in a small band dispersion in the interlayer direction ($\Gamma$--A path), which is more noticeable in the iodide than in the chloride and bromide as expected from the degree of proximity of the halogen-$p$ energy levels. 

The band inversion observed in the above hypothetical structures implies an interesting possibility in the chloride system, whose properties have been less explored before. Since the low-energy valence bands at $\Gamma$ are nearly degenerate in the chloride, by exerting a uniaxial tensile strain on this, the difference $d_{2}^{{\rm Y-Y}}\!-d_{4}^{{\rm Y-Y}}$ is enlarged and a non-monotonic change in its electronic characteristics may be observed due to the band inversion. The same would also be true for the bromide system, though the required strain should be stronger.

\begin{figure*}[t]
 \begin{center}
  \includegraphics[scale=0.35]{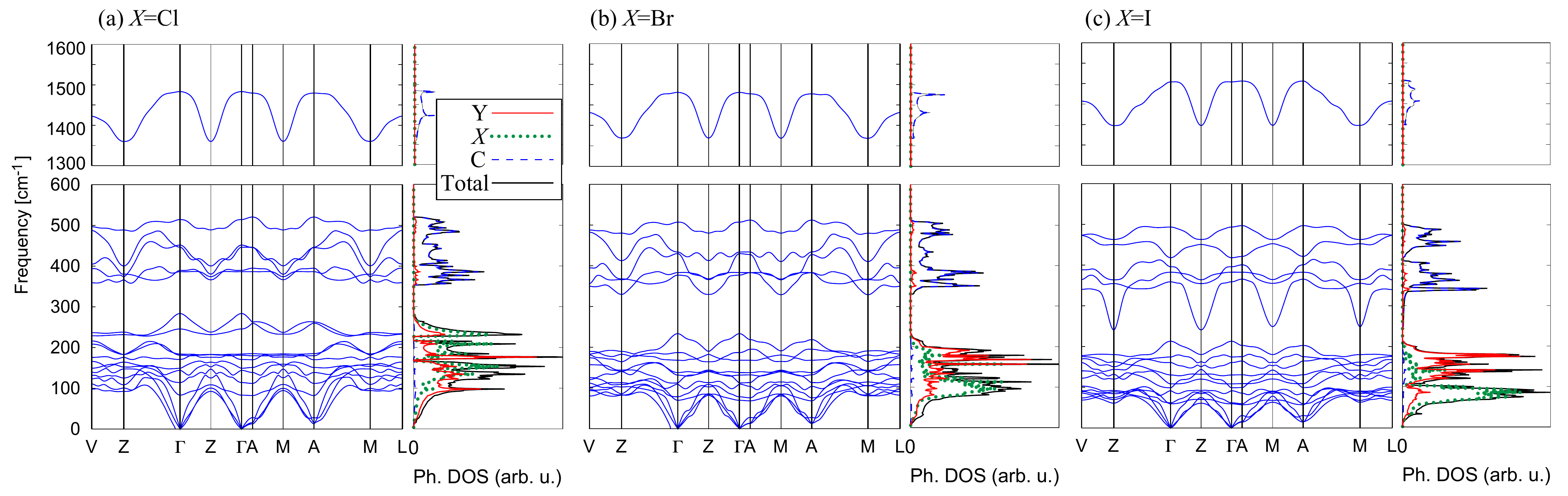}
  \caption{Phonon dispersion and DOS of Y$_{2}$$X_{2}$C$_{2}$. The components of the phonon DOS projected onto the respective atoms are also shown.} 
  \label{fig:ph-sum}
 \end{center}
\end{figure*}

\begin{figure}[t]
 \begin{center}
  \includegraphics[scale=0.35]{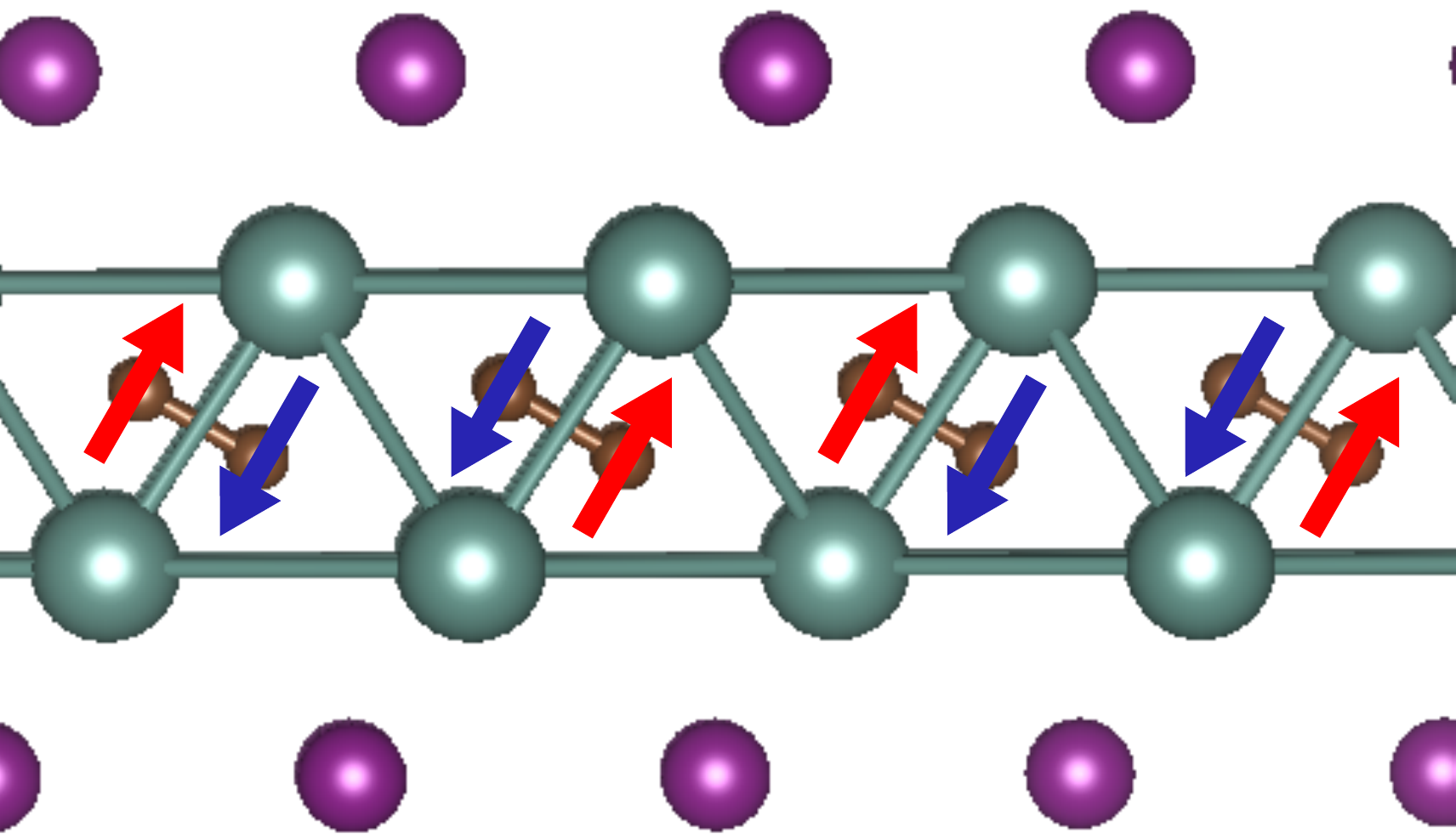}
  \caption{Side view of the softened carbon dimer libration mode at Z and M in Y$_{2}$I$_{2}$C$_{2}$, where the mode Bloch phase in the layer direction is reflected.} 
  \label{fig:soft-mode}
 \end{center}
\end{figure}

\subsection{Phononic properties}

\begin{table}[t]
\caption[t]
{Calculated phonon frequencies at $\Gamma$. The reference experimental and first-principles calculation data are taken from Refs.~\onlinecite{Henn-neutron-PRB1998} and ~\onlinecite{Puschnig-PRB2001}, respectively.}
\begin{center}
\label{tab:phonon-modes}
\tabcolsep = 1mm
\begin{tabular}{lccccccc} \hline
\multicolumn{1}{c}{  } &\multicolumn{1}{c}{$X=$Cl} & \multicolumn{3}{c}{$X=$Br} & \multicolumn{3}{c}{$X=$I}  \\ 
  (cm$^{-1}$)& Present & Present & Expt. &Calc. &Present & Expt. &Calc. \\ \hline
$A_{\rm g}$ & 1483 &1483 &1593&1589 &1503&1588&1527  \\
$A_{\rm g}$  & 451 & 432 &436&436 &398&401&403  \\
$A_{\rm g}$  & 284 &233 &236&232 &213&215 &219 \\ 
$A_{\rm g}$  & 172 &  169&166&172 &174&169 &173 \\
$A_{\rm g}$  & 144 &109 &108&113 &91&93 &94 \\ 
$A_{\rm g}$  & 114&79  & &79&62& &59 \\
$B_{\rm g}$  & 446 & 436& &436&420& &421 \\
$B_{\rm g}$  & 181 & 138&140 &131&130&126 &128 \\
$B_{\rm g}$  & 83&70  &67 &66&59&58 & 58\\    \hline
$A_{\rm u}$  & 386 & 386& & & 384& & \\
$A_{\rm u}$  & 184 & 132& & & 109& & \\
$B_{\rm u}$  & 514 & 513& & & 491& & \\
$B_{\rm u}$  & 388 & 379& & & 361& & \\
$B_{\rm u}$  & 246 & 187& & &161 & &  \\
$B_{\rm u}$  & 149 & 114& & & 98& & \\ \hline
\end{tabular}
\end{center}
\end{table}

We next examine the calculated phonon spectra shown in Fig.~\ref{fig:ph-sum}. The branches form three distinct groups; the low-frequency branches (less than 300~cm$^{-1}$), medium-range frequency branches (between 250~cm$^{-1}$ and 500~cm$^{-1}$) and the single high-frequency mode. We refer to these branches and corresponding frequency ranges as ``low", ``medium", and ``high" in the later discussions, respectively. From the analysis of the phonon DOS (right panels) and mode vectors, the lowest branches are dominated by the yttrium and halogen atoms whereas the medium-frequency branches correspond to the motions of the carbon dimers as a rigid body (three translations and two librations). The C-C stretch of the $C_{2}$ dimer has the highest frequency. These are consistent with the previous experimental characterization of the Raman-active modes~\cite{Henn-neutron-PRB1998} and first-principles simulation.\cite{Puschnig-PRB2001} Note that the appearance of the dimer rigid-body vibrations at intermediate-frequency regimes is a feature common to YC$_{2}$ (Ref.~\onlinecite{XUE2019120}).

We summarize the frequencies of the calculated Raman and infrared-active modes in Table~\ref{tab:phonon-modes}. Except for the carbon dimer stretching mode, the present theoretical predictions agree well with the preceding experimental and first-principles results. The underestimation for the stretching mode could be due to the use of the PBEsol exchange-correlation functional, of which the exchange part is made more accurate in solid systems rather than in molecules.~\cite{PBEsol}

In the calculated phonon spectra in Fig.~\ref{fig:ph-sum}, we observe two significant halogen dependences. First, the halogen-atom component of the projected vibrational DOS shifts downward in frequency with replacement of the lighter halogen by heavier atoms, which is a trivial consequence of the change in the atomic mass. Second, more strikingly, one of the carbon modes shows significant softening at the $Z$ and $M$ points by the halogen-atom substitution. This mode was found to be a collective libration of the carbon dimers as depicted in Fig.~\ref{fig:soft-mode}. This softening implies that the 1$s$-Y$_{2}$I$_{2}$C$_{2}$ having the highest $T_{\rm c}$ is on the verge of some structural instability related to this libration, which is apparently distinct from the 1$s$-3$s$ transition under pressure.~\cite{Ahn-pressure-JPhysC2016}


\begin{figure}[t]
 \begin{center}
  \includegraphics[scale=0.50]{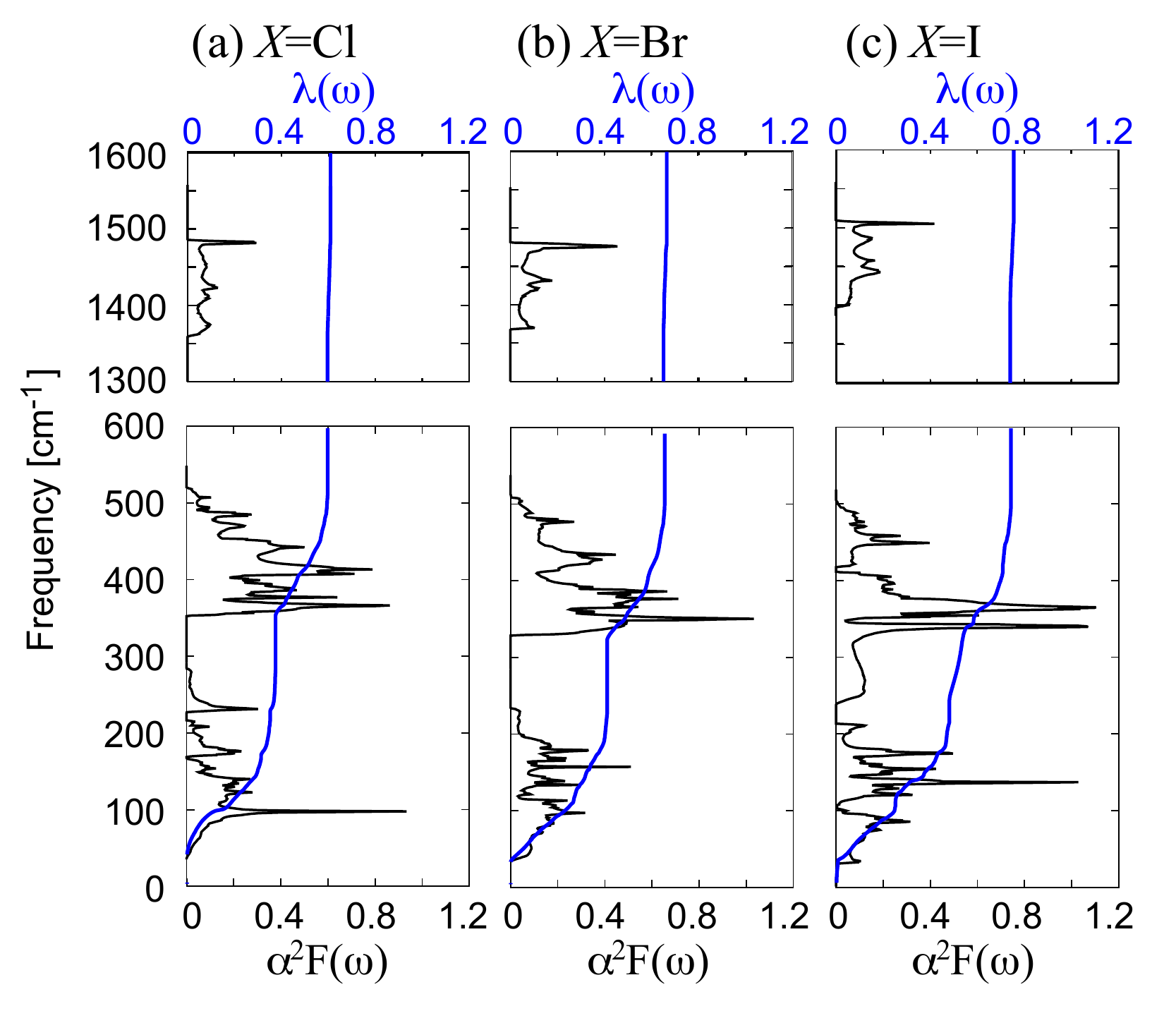}
  \caption{Eliashberg spectral function $\alpha^{2}F$ and partially integrated $\lambda(\omega)$ defined in Eq.~(\ref{eq:partial-lambda}) for Y$_{2}$$X_{2}$C$_{2}$.} 
  \label{fig:a2F-lambda-sum}
 \end{center}
\end{figure}


\begin{figure}[h]
 \begin{center}
  \includegraphics[scale=0.50]{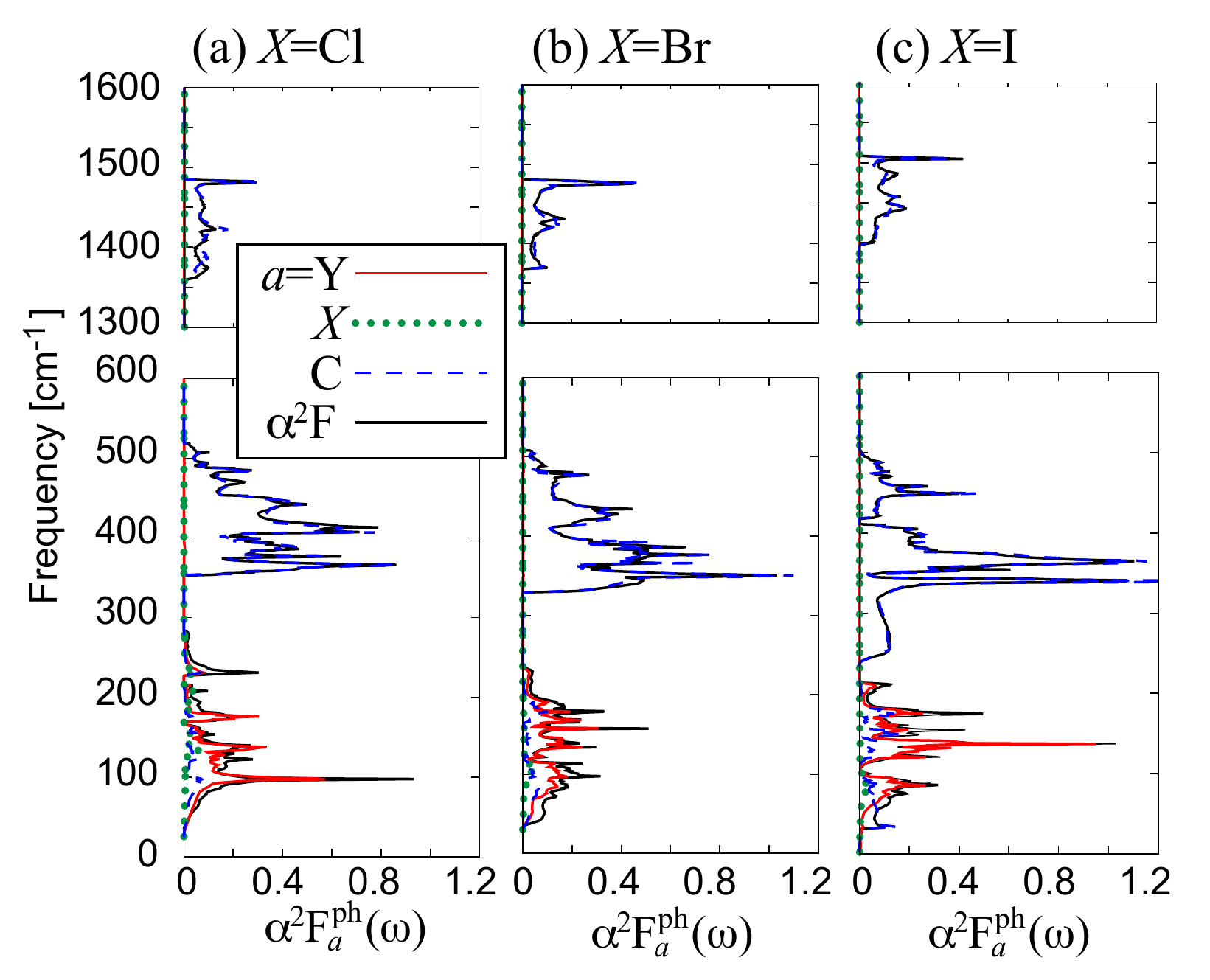}
  \caption{Partially summed components of the Eliashberg function defined in Eq.~(\ref{eq:a2F-partial-ph}).} 
  \label{fig:a2F-decomp-ph}
 \end{center}
\end{figure}

\begin{figure}[h]
 \begin{center}
  \includegraphics[scale=0.50]{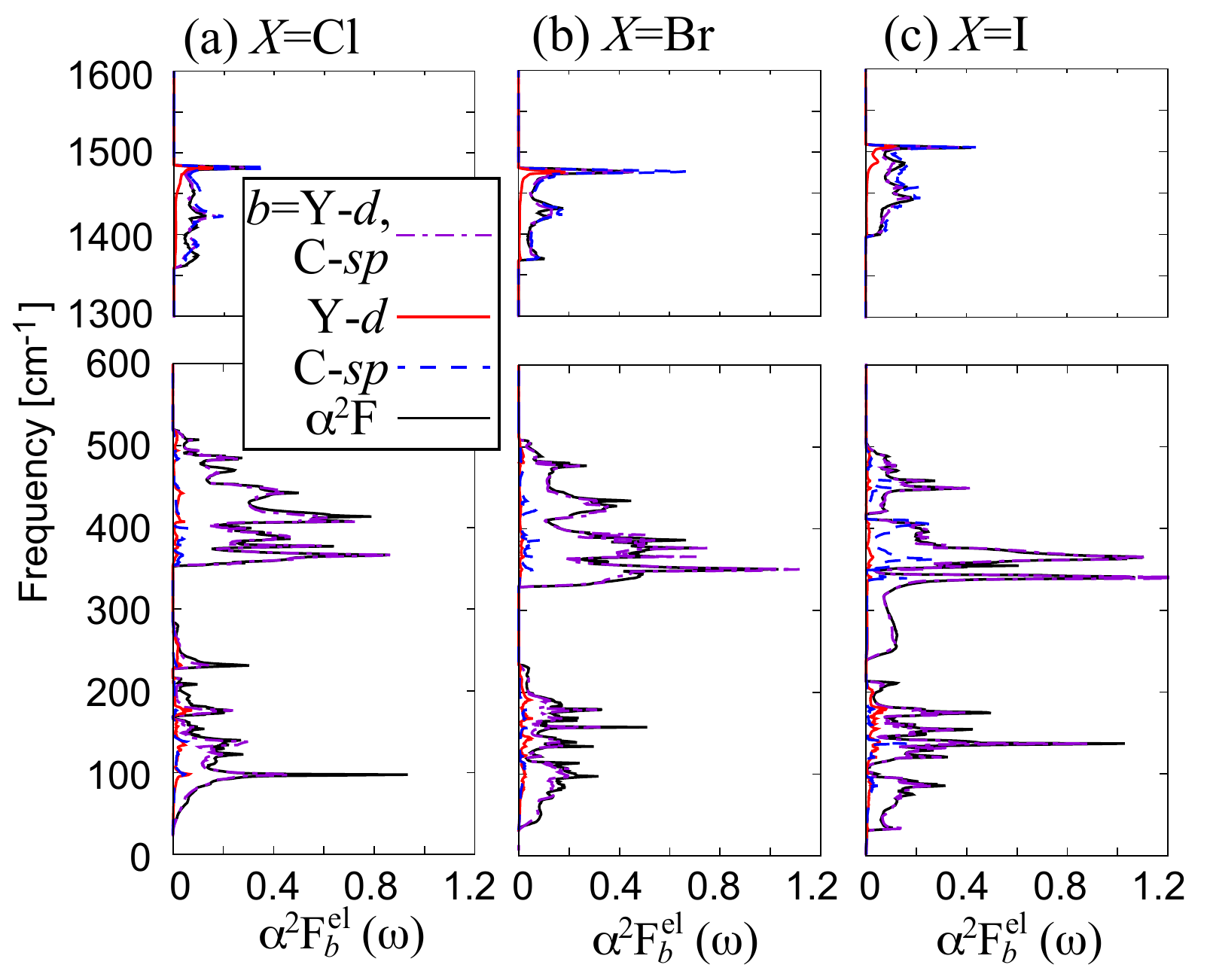}
  \caption{Partially summed components of the Eliashberg function defined in Eq.~(\ref{eq:a2F-partial-el}).} 
  \label{fig:a2F-decomp-el}
 \end{center}
\end{figure}

\subsection{Electron-phonon coupling properties and superconducting $T_{\rm c}$}

We show in Fig.~\ref{fig:a2F-lambda-sum} the calculated Eliashberg spectral function $\alpha^{2} F(\omega)$ and partially integrated $\lambda(\omega)$, which converges to $\lambda$ [Eq.~(\ref{eq:lambda})] in the $\omega \rightarrow \infty$ limit, defined as
\begin{eqnarray}
\lambda(\omega)
=
2\int _{0}^{\omega} d\omega' \frac{\alpha^{2} F(\omega')}{\omega'}
.
\label{eq:partial-lambda}
\end{eqnarray}
For all the systems, approximately half of the total $\lambda$ originates from the coupling of the lowest yttrium-halogen branches. Most of the remainder is from the translation and libration modes of the carbon dimers, whereas the contribution of the stretching mode is little.  The large fraction of the yttrium-halogen contribution explains the not so appreciable isotope effect found in experiments.~\cite{Schnelle-isotope-JAP1998} The halogen dependence of the total $\lambda$ (Table~\ref{tab:elph-Tc}) agrees well with the experimentally observed trend of $T_{\rm c}$, with which we can actually reproduce the values of $T_{\rm c}$ from first principles in the right order as seen later. As the halogen atoms become heavier, the characteristic frequencies $\omega_{\rm ln}$ and $\omega_{2}$ monotonically decrease. The mechanism of this seems simply understandable from the change of the atomic mass, but below we find that the situation is not that obvious. The area $A$ is almost constant. The relations of $\omega_{\rm ln}$, $\omega_{2}$ and $A$ to $T_{\rm c}$ are discussed below.

Here we elucidate the origin of the coupling with the local decompositions $\alpha^{2}F^{\kappa\kappa'}_{m_{1}m_{2} m_{3}m_{4}} (\omega)$ defined in Eq.~(\ref{eq:a2F-decomp}). We first define a partial sum of them as
\begin{eqnarray}
\alpha^{2}F^{\rm ph}_{a}
=
\sum_{\kappa, \kappa' \in a}\sum_{\substack{m_{1}m_{2} \\ m_{3}m_{4}}}\alpha^{2}F^{\kappa\kappa'}_{m_{1}m_{2} m_{3}m_{4}} (\omega)
.
\label{eq:a2F-partial-ph}
\end{eqnarray}
The operation $\sum_{\kappa, \kappa' \in a}$ denotes the partial sum for the displacement $\kappa$ and $\kappa'$ corresponding to the atomic kind $a$ (=Y, $X$, C). This function indicates the partial contribution to $\alpha^{2}F(\omega)$ of the vibrations of the specific atoms. Similarly, we also define another partial sum as
\begin{eqnarray}
\alpha^{2}F^{\rm el}_{b}
=
\sum_{m_{1}m_{2}m_{3}m_{4} \in b}\sum_{\kappa, \kappa' }\alpha^{2}F^{\kappa\kappa'}_{m_{1}m_{2} m_{3}m_{4}} (\omega)
.
\label{eq:a2F-partial-el}
\end{eqnarray}
The sum $\sum_{m_{1}m_{2}m_{3}m_{4} \in b}$ runs over the specific group of Wannier orbitals $b$ and extracts the contributions of the target Wannier orbitals. The corresponding partial contributions to $\lambda$ are defined as $\lambda^{\rm ph}_{a}$ and $\lambda^{\rm el}_{b}$, respectively.

\begin{figure}[h]
 \begin{center}
  \includegraphics[scale=0.25]{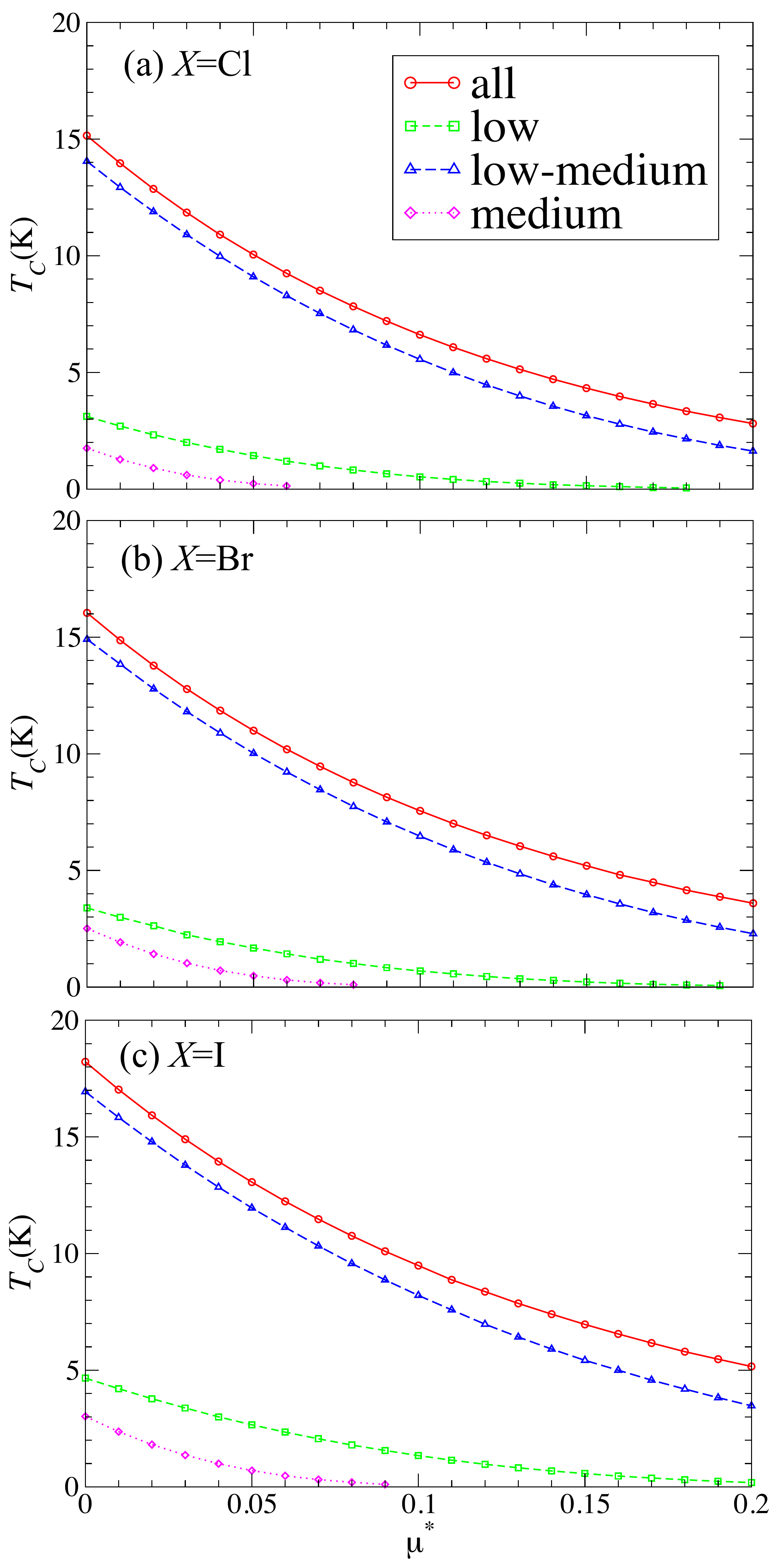}
  \caption{Transition temperatures calculated by the Eliashberg equations with varying Coulomb repulsion parameter $\mu^{\ast}$. Legends indicate the frequency ranges where input $\alpha^2 F(\omega)$ is kept to nonzero.}
  \label{fig:Tc-Eliashberg}
 \end{center}
\end{figure}

Figure~\ref{fig:a2F-decomp-ph} displays $\alpha^{2}F^{\rm ph}_{a}$ for $a=$Y, $X$, C. An important finding is that the component $a=X$ is negligibly small. The low-frequency branches are formed by entangled vibrations of yttrium and halogen atoms and from the total $\alpha^{2}F$ we cannot see which kind of atoms mainly contribute: Our results clarify that the electron scattering processes due to the vibrations of halogen atoms are irrelevant. The partial $\lambda^{\rm ph}$ in Table~\ref{tab:elph-Tc} indicates that the contributions of the yttrium and carbon vibrations are of equal importance. Here, the reduction of $\omega_{\rm ln}$ and $\omega_{2}$ by the halogen substitution from Cl, Br to I (Table~\ref{tab:elph-Tc}) is revealed to be an indirect effect which cannot be attributed to the coupling of the halogen-atom vibration. As mentioned in Sec.~\ref{sec:struct}, the substitution of the halogen atoms induces the uniaxial compression of the Y$_{2}$C$_{2}$ blocks. This results in softening of the C$_{2}$ libration modes strongly coupled to the conducting electrons and downward shift of the medium-range frequency component of $\alpha^{2}F(\omega)$.

In Fig.~\ref{fig:a2F-decomp-el}, we show $\alpha^{2}F^{\rm el}_{b}$ for $b=$\{Y-$d$, C-$sp$\}, \{Y-$d$\} and \{C-$sp$\}. The components involving the $X$-$sp$ orbitals are presumably negligible according to the small projected DOS in Fig.~\ref{fig:bands-sum}. We note that the $b=$\{Y-$d$, C-$sp$\} component reproduces the total $\alpha^{2}F$ almost completely, whereas the $b=$\{Y-$d$\} and $b=$\{C-$sp$\} components are much smaller. In the latter two, electron scattering processes across the Y-$d$ and C-$sp$ orbitals are ignored. These ignored processes are thus found to be essential for reproducing the total $\alpha^{2}F$ and, in particular, the small $b=$\{C-$sp$\} component directly demonstrates that the Zintl-Klemm pairing theory, or the ``C$_2$ units as pairing traps" scenario of Simon~\cite{Simon-review} does not simply apply.


We have thus clarified origins of the three major branches of the $\alpha^2 F(\omega)$ spectra. To understand their effects on $T_{\rm c}$, we solved the transformed Eliashberg equation Eq.~(\ref{eq:Iso-Eliashberg3}) and calculated $T_{\rm c}$ with modifications to the input $\alpha^2 F(\omega)$ so that it is suppressed to zero at certain frequency ranges. The values of $T_{\rm c}$ with varying the empirical Coulomb parameter $\mu^{\ast}$ are shown in Fig.~\ref{fig:Tc-Eliashberg}. By suppressing its amplitude to zero for the high-frequency branch, the resulting $T_{\rm c}$ values are lowered only by $\sim$1~K (``all" to ``low-medium"), which indicates the small coupling effect of the carbon stretching modes. If we further suppress either of the low and medium frequency branches (from ``low-medium" to ``medium" or ``low"), $T_{\rm c}$ significantly decreases. The latter result shows that the origin of the observed $T_{\rm c}$s cannot be attributed solely to either of the carbon rigid modes or yttrium modes.

In the above calculations, the values of $T_{\rm c}$ were qualitative estimates because of the undetermined empirical value $\mu^{\ast}$. We finally solved the SCDFT gap equation [Eq.~(\ref{eq:SCDFTgap})] using the calculated $\alpha^{2}F$ and other electronic normal-state properties and evaluated $T_{\rm c}$s nonempirically. The resulting values in Table~\ref{tab:elph-Tc} reproduce the experimentally observed halogen dependence: $T_{\rm c}(X=$Cl$)<T_{\rm c}(X=$Br$)<T_{\rm c}(X=$I$)$. This result indicates that the Migdal-Eliashberg picture is valid for describing superconductivity in this series of compounds. We note that $T_{\rm c}$ of Y$_{2}$I$_{2}$C$_{2}$ is significantly underestimated. The discrepancy may be a hint that physics beyond the Migdal-Eliashberg theory should become relevant in this system. Here we stress a possible importance of the softened C$_{2}$ libration mode in Fig.~\ref{fig:soft-mode}. Despite the small phonon DOS seen in Fig.~\ref{fig:ph-sum}, this branch has contribution by $\sim$ 0.1 to the total $\lambda$. It is hence inferred that the electron scattering strength of the libration mode [represented by $\alpha^{2}F$/(phonon DOS)] is anomalously large. Higher-order vertex corrections involving this mode can have appreciable effect or, more drastically, charge density wave fluctuations induced by the strong interaction might partially break the conventional pairing picture.

\begin{figure}[h]
 \begin{center}
  \includegraphics[scale=0.25]{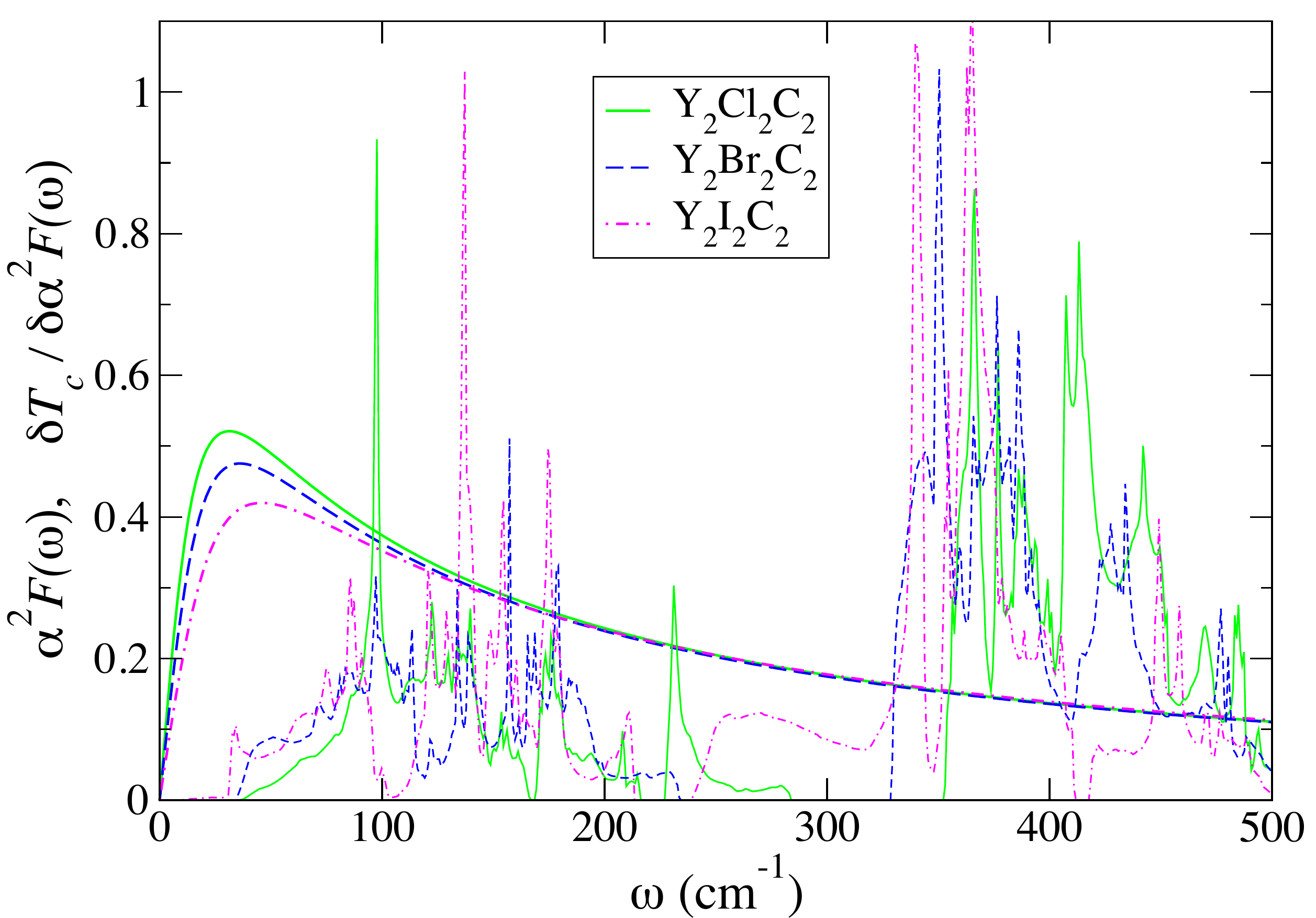}
  \caption{Functional derivative of $T_{\rm c}$ with respect to $\alpha^2 F(\omega)$ with the $\alpha^2 F(\omega)$ spectra. The empirical parameter $\mu^{\ast}$ was set to 0.13.} 
  \label{fig:dTc-da2F}
 \end{center}
\end{figure}

The halogen dependence of the calculated $T_{\rm c}$ well conforms to that derived from the semiempirical McMillan-Allen-Dynes formula $T_{\rm c}^{\rm McM}$ [(Eq.~\ref{eq:McMillan}); Table~\ref{tab:elph-Tc}]. We can therefore interpret the origin of the increase of $T_{\rm c}$ upon the halogen substitution with the parameters entering the formula. In terms of $\lambda$, $\omega_{\rm ln}$ and $\omega_{2}$, it is due to the enhancement of $\lambda$ dominating over the reduction of $\omega_{\rm ln}$ and $\omega_{2}$. Another interpretation is derived by focusing on the almost constant area $A$. We show in Fig.~\ref{fig:dTc-da2F} the calculated functional derivative $\delta T_{\rm c}/\delta \alpha^{2}F(\omega)$ for $T_{\rm c}$ derived from the Eliashberg equation. The functional derivative takes its maximum value for all the three systems at $\lesssim$50 cm$^{-1}$$\simeq$ 70~K, which is consistent with the preceding studies where the maximum was observed at $\omega \sim 7 T_{\rm c}$.~\cite{Bergmann-Rainer, Karakozov-SovPhys, Carbotte-RMP1990} With $A$ kept constant, $T_{\rm c}$ is therefore raised by concentration of $\alpha^{2}F(\omega)$ around 7$T_{\rm c}$ $\sim$ 50~cm$^{-1}$. In this respect, the smaller $\omega_{\rm ln}$ and $\omega_{2}$ are more advantageous for $T_{\rm c}$ in Y$_{2}$$X_{2}$C$_{2}$, which is why $T_{\rm c}$ increases despite the decrease of those frequencies. In other words, Y$_{2}$I$_{2}$C$_{2}$ hosts the optimum distribution of $\alpha^{2}F$ among the halides with a given total area under it.

Although we have reproduced the proper halogen dependence of $T_{\rm c}$, how it is induced via the distortion of the Y$_{2}$C$_{2}$ blocks is not as simple according to the present calculations. The electronic DOS at the Fermi energy $N(0)$, to which the total coupling $\lambda$ is approximately proportional, shows a slight increase by halogen substitution from the lighter to heavier atoms as pointed out in the previous first principles band structure calculation.\cite{Puschnig-PRB2001} However, it does not completely explain the enhancement of the total $\lambda$. Thus, in this work, we stop with the remark that the distortion of the Y$_{2}$C$_{2}$ blocks induces the enhancement of the electron-phonon matrix elements as well as the slight increase of the DOS. More detailed theory on its mechanism is left for future studies.

\begin{table}[t]
\caption[t]
{Parameters representing the phonon mediated pairing and superconducting transition temperatures from first principles. See text for the definitions of $\lambda^{\rm ph}$.}
\begin{center}
\label{tab:elph-Tc}
\tabcolsep = 1mm
\begin{tabular}{l ccc} \hline
&YClC & YBrC & YIC  \\ \hline
$N(0)$ [(eV sp)$^{-1}$] & 0.781 & 0.803  & 0.859  \\ 
$\lambda$ & 0.610 & 0.667  & 0.755  \\ 
$\lambda_{\rm Y}^{\rm ph}$ & 0.288 & 0.271  & 0.319  \\ 
$\lambda_{\rm C}^{\rm ph}$ & 0.314 & 0.360  & 0.484  \\  \hline
$\omega_{\rm ln}$ (K) & 273 & 244 & 224  \\ 
$\omega_{2}$ (K) & 489 & 468 & 448  \\
$A$ (K) & 111 & 114 & 119  \\ \hline
$T_{\rm c}^{\rm McM}$ ($\mu^{\ast}$$=$$0.13$) (K)&4.7  &5.6  &7.3   \\ 
$T_{\rm c}^{\rm SCDFT}$ (K)& 2.6 &3.3 &4.6   \\ 
$T_{\rm c}^{\rm exp}$ (K) &2.3  &5.0  &10.0   \\ \hline
\end{tabular}
\end{center}
\end{table}

\section{Discussions and summary}

Through first-principles calculations of the structural and electron-phonon properties, we have established an understanding on superconductivity in the yttrium carbide halide. The Wannier-based analyses of the electron-phonon coupling have revealed that the mechanism of the $T_{\rm c}$ change by the halogen substitution is rather indirect. The halogen ions affect the pairing solely via the local uniaxial compressive force distorting the Y$_{2}$C$_{2}$ blocks of the Poisson type, which induces changes in the low-energy electronic band structure, phonon spectra of the Y-$X$ entangled vibrations, and electron-phonon coupling character in the blocks. Neither their electronic orbitals nor ionic displacements directly participate in the superconducting pairing mechanism. Within the superconducting Y$_{2}$C$_{2}$ blocks, rigid-body vibrations of the carbon dimers and vibrations of the yttrium frameworks containing them are energetically well decoupled, but both have appreciable contributions to the total pairing. We cannot attribute the pairing electrons to either Y-$d$ or C-$sp$ electronic orbitals since the scattering processes across them are essential. In that sense, the Y-$d$ or C-$sp$ orbitals are entangled from the electron-phonon coupling aspect. The simple perspective with emphasis on the C$_{2}$ molecular orbitals is hence invalid. The coupling of the C$_{2}$ bond stretching mode is, contrary to the expectation, small and have little effect on $T_{\rm c}$: Its values of order 1--10~K is due to the moderately high frequencies of the C$_{2}$ rigid-body modes interacting with both the Y-$d$ and C-$sp$ orbitals. The irrelevance and relevance of those C$_{2}$ modes are common to yttrium sesquecarbide Y$_{2}$C$_{3}$  (Ref.~\onlinecite{Singh-Mazin-Y2C3}).

According to the calculated values of $T_{\rm c}$ and electron-phonon coupling parameters, the halogen-atom dependence of $T_{\rm c}$ is due to the difference of the electron-phonon matrix elements within the Y$_{2}$C$_{2}$ blocks, not only of the electronic DOS at the Fermi energy, although direct explanation as to how they occur upon the structural Poisson effect is left for future studies. In the most distorted iodide system, anomalous frequency softening of the collective carbon libration mode has been observed, with which the calculated $T_{\rm c}$ with the phonon-mediated SCDFT substantially departs from the experimentally observed value. Because this softening is accompanied by the enhancement of the corresponding electron-phonon coupling matrix element, it may have effects beyond the Migdal-Eliashberg picture. The Y$_{2}$I$_{2}$C$_{2}$ system hence serves as a testbed for quantitative numerical methods for unconventional mechanisms. Recently the quantitative difference between the non-empirical Eliashberg theory~\cite{Sanna2018} and SCDFT has been examined.~\cite{SPG2020} A possibility arises that improvement of the SCDFT kernels for the phonon pairing~\cite{SPG2020} may correct the current underestimation of $T_{\rm c}$ for the bromide and iodide to some extent: It is also an important issue for testing the applicability of the conventional phonon-mediated mechanism.


Apart from the mechanism, the remarkable correlation between the Poisson distortion of the Y$_{2}$C$_{2}$ blocks and $T_{\rm c}$ suggests an interesting path to higher $T_{\rm c}$. Namely, uniaxial compression of the system may enhance the distortion of the yttrium frameworks. In a previous experiment, hydrostatic pressure has been exerted on Y$_{2}$I$_{2}$C$_{2}$ (Refs.~\onlinecite{Ahn2005, Ahn-pressure-JPhysC2016}). There, enhancement of $T_{\rm c}$ up to 11.5~K was observed at the critical pressure where the stacking changes from $1s$ to $3s$. Similar enhancement has been observed with partial halogen substitution in Y$_{2}$(Br, I)$_2$C$_{2}$ (Refs.~\onlinecite{Simon-ZAAC1996} and \onlinecite{Henn-BrI-2001}). They probably originate from the structural instability on the verge of the $1s$-$3s$ phase transformation. The uniaxial pressure, which is expected to further soften the carbon libration mode, may drive the system toward another structural instability. The carbon libration modes should then be ``mixed (Ref.~\onlinecite{Tse-PRB2017})" with the yttrium modes and the distribution of $\alpha^{2}F(\omega)$ becomes more concentrated around $\sim$7$T_{\rm c}$, close to the optimal condition for $T_{\rm c}$.

Lastly, we note the potential generality of the concept of internal structural compression generated by the halogen atoms. Effect of replacing the atoms within the same group is interpreted as the change of the radii of the ions at the corresponding positions. The point is that, in compounds containing two or more kinds of atoms, the stress induced by varying the atomic radii is generally nonuniform. It can therefore induce expansion of a certain structure and compression of another simultaneously. In the current Y$_{2}$$X_{2}$C$_{2}$ system (and probably in other layered systems), the replacement of the intercalant ions was found to induce the uniaxial compression and lateral expansion of the superconducting blocks, which cannot be directly inferred from the change of the monotonically increasing lattice parameters. More generally one would find a variety of structural deformations induced by the nonuniform electronic stress, which will provide us with a novel method of controlling the properties of materials in addition to exerting external pressures.

\section*{Acknowledgment}
R.~Ak. thanks to Mitsuaki Kawamura and Yusuke Nomura for providing subroutines for the tetrahedron integrations of the Eliashberg function and calculation of the electron-phonon matrix elements in the Wannier gauge, respectively. This work was supported by JSPS KAKENHI Grant Numbers 15K20940 (R. Ak.) and 19H05825 (R. Ar.) from Japan Society for the Promotion of Science (JSPS) and by Natural Sciences and Engineering Research Council of Canada. C. Z. acknowledges the financial support from the National Natural Science Foundation of China (Grant No. 11874318) and the Natural Science Foundation of Shandong Province (Grant No. ZR2018MA043).  J. S. T. wishes to thank Compute Canada for allocation of computing resources. Some calculations were performed at the Supercomputer Center at the Institute for Solid State Physics in the University of Tokyo and the SGI Rackable C2112-4GP3/C1102-GP8 (Reedbush-U/H/L) in the Information Technology Center, The University of Tokyo.

\bibliography{reference}

\end{document}